\documentclass{article}
\usepackage[utf8]{inputenc}
\usepackage[T1]{fontenc}
\usepackage{hyperref}
\usepackage{url}
\usepackage{booktabs}
\usepackage{amsmath, amssymb, amsfonts}
\usepackage{algorithm}
\usepackage{algorithmic}
\usepackage{microtype}
\usepackage{url}
\usepackage{graphicx}
\usepackage[numbers]{natbib}
\usepackage{cite}
\usepackage{xcolor}
\usepackage{booktabs}
\usepackage{siunitx}
\usepackage{float}      
\usepackage{placeins}   
\usepackage[dblblindworkshop, final]{neurips_2025}
\title{Lark: Biologically Inspired Neuroevolution for Multi-Stakeholder LLM Agents}
\workshoptitle{Workshop on Efficient Reasoning}

\author{
Rikhil Tanugula \\
Aelin\\
San Jose, CA \\
\texttt{rikhil@aelin.tech} \\
\And
Dheeraj Chintapalli \\
Reteena\\
Las Vegas, NV \\
\texttt{dheeraj@reteena.org} \\
\And
Sunkalp Chandra \\
Reteena\\
Manalapan, NJ \\
\texttt{sunkalp@reteena.org} \\
}

\begin{document}
\maketitle

\begin{abstract}
We present Lark, a biologically inspired decision-making framework that couples LLM-driven reasoning with an evolutionary, stakeholder-aware Multi-Agent System (MAS). To address verbosity and stakeholder trade-offs, we integrate four mechanisms: (i) plasticity, which applies concise, context-sensitive adjustments to candidate solutions; (ii) duplication and maturation, which copy high-performing candidates and specialize them into new modules; (iii) ranked-choice stakeholder aggregation using influence-weighted Borda scoring; and (iv) compute awareness via token-based penalties that reward brevity and update an efficiency metric each generation. The system iteratively proposes diverse strategies, applies plasticity tweaks, simulates stakeholder evaluations, aggregates preferences, selects top candidates, and performs duplication/maturation while factoring compute cost into final scores. In a controlled evaluation over 30 rounds comparing 14 systems, Lark Full achieves a mean rank of \(2.55\) (95\% CI \([2.17, 2.93]\)) and a mean composite score of \(29.4/50\) (95\% CI \([26.34, 32.46]\)), finishing Top-3 in 80\% of rounds while remaining cost competitive with the leading commercial models (\$0.016 per task). Paired Wilcoxon tests confirm that all four mechanisms contribute significantly as ablating duplication/maturation yields the largest deficit (\(\Delta\)Score \(= 3.5\), Cohen's \(d_z = 2.53\), \(p < 0.001\)), followed by plasticity (\(\Delta\)Score \(= 3.4\), \(d_z = 1.86\)), ranked-choice voting (\(\Delta\)Score \(= 2.4\), \(d_z = 1.20\)), and token penalties (\(\Delta\)Score \(= 2.2\), \(d_z = 1.63\)). Rather than a formal Markov Decision Process (MDP) with constrained optimization, Lark is a practical, compute-aware neuroevolutionary loop that scales stakeholder-aligned strategy generation and makes its trade-offs transparent through per-step metrics and final analyses. Our work presents proof-of-concept findings and invites community feedback as we expand toward real-world validation studies.
\end{abstract}

\section{Introduction}
Multi-agent systems (MAS) play a crucial role in AI tasks ranging from collaborative robotics, social simulations, economics, autonomous vehicles, and healthcare planning~\citep{Wooldridge2009, Lim2024, Vineis2025}. Traditional MAS are effective in optimizing fixed objectives but fail in domains requiring creative strategy generation, trade-off evaluation, and alignment with multiple, often conflicting stakeholder objectives~\citep{Radulescu2021,Han2024}.

Large language models (LLMs) provide new capabilities for MAS by enabling agents to reason abstractly, simulate human stakeholders, and generate novel strategies~\citep{Tran2025, SultanaChowa2025}. However, current LLM-augmented MAS face three limitations:
\begin{enumerate}
    \item Verbose and redundant outputs leading to high computational costs.
    \item Poor exploration of alternative strategies and potential future states.
    \item Inadequate mechanisms for multi-stakeholder and multi-objective decision-making.
\end{enumerate}

\textbf{Lark} addresses these challenges by integrating four biologically inspired mechanisms:
\begin{itemize}
    \item \textbf{Plasticity:} synaptic weights adapt within an agent’s lifetime in response to context.
    \item \textbf{Duplication and maturation:} high-performing modules are copied and specialized to improve diversity and modularity.
    \item \textbf{Stakeholder-ranked voting:} multi-objective aggregation using Borda count.
    \item \textbf{Compute-efficiency penalties:} explicit constraints encourage concise reasoning and sparse architectures.
\end{itemize}

We formalize Lark as a discrete, compute-aware \emph{evolutionary} search over strategy populations, rather than an MDP, where strategies are generated and evaluated holistically, not via stepwise state-action transitions. We also conduct comprehensive experiments on synthetic benchmarks demonstrating adaptation, fairness, and modularity improvements.

\section{Related Work}
\subsection{Multi-Agent Systems and LLMs}
MAS research traditionally focuses on collaborative optimization using reinforcement learning, evolutionary strategies, or planning~\citep{Silver2016, Clune2013, Vineis2025}. Systems such as AlphaGo~\citep{Silver2016} achieve superhuman performance in single-objective tasks but do not support multi-objective optimization, intra-lifetime adaptation, or compute efficiency constraints.

LLM-augmented MAS enhance reasoning and persistent memory~\citep{Lim2024, Tran2025}, allowing for strategy simulation, stakeholder modeling, and creative planning. Despite these advances, they lack structured mechanisms for efficient multi-objective exploration.

\subsection{Biologically Inspired Neuroevolution}
Neuroevolution exploits biological principles such as plasticity, duplication, modularity, and epigenetics to improve adaptation~\citep{Soltoggio2008, Stanley2002, Mouret2012}. Mechanisms include:
\begin{itemize}
    \item \textbf{Plasticity:} dynamic weight updates within a lifetime allow rapid adaptation~\citep{Fang2017}.
    \item \textbf{Duplication and maturation:} modular components are duplicated and specialized, promoting diversity and reusability~\citep{Stanley2002}.
    \item \textbf{Epigenetic adaptation:} modifications beyond genetic encoding enable agents to respond to contextual changes~\citep{Yuen2023, Kriuk2025}.
    \item \textbf{Socially inspired aggregation:} multi-objective ranking and voting mechanisms enhance collective decision-making~\citep{Black1986, Vineis2025}.
\end{itemize}

No prior work combines all four mechanisms in LLM-driven MAS with explicit compute-aware constraints.

\section{Methodology}

\begin{figure}[htbp]
\centering
\includegraphics[width=0.8\linewidth]{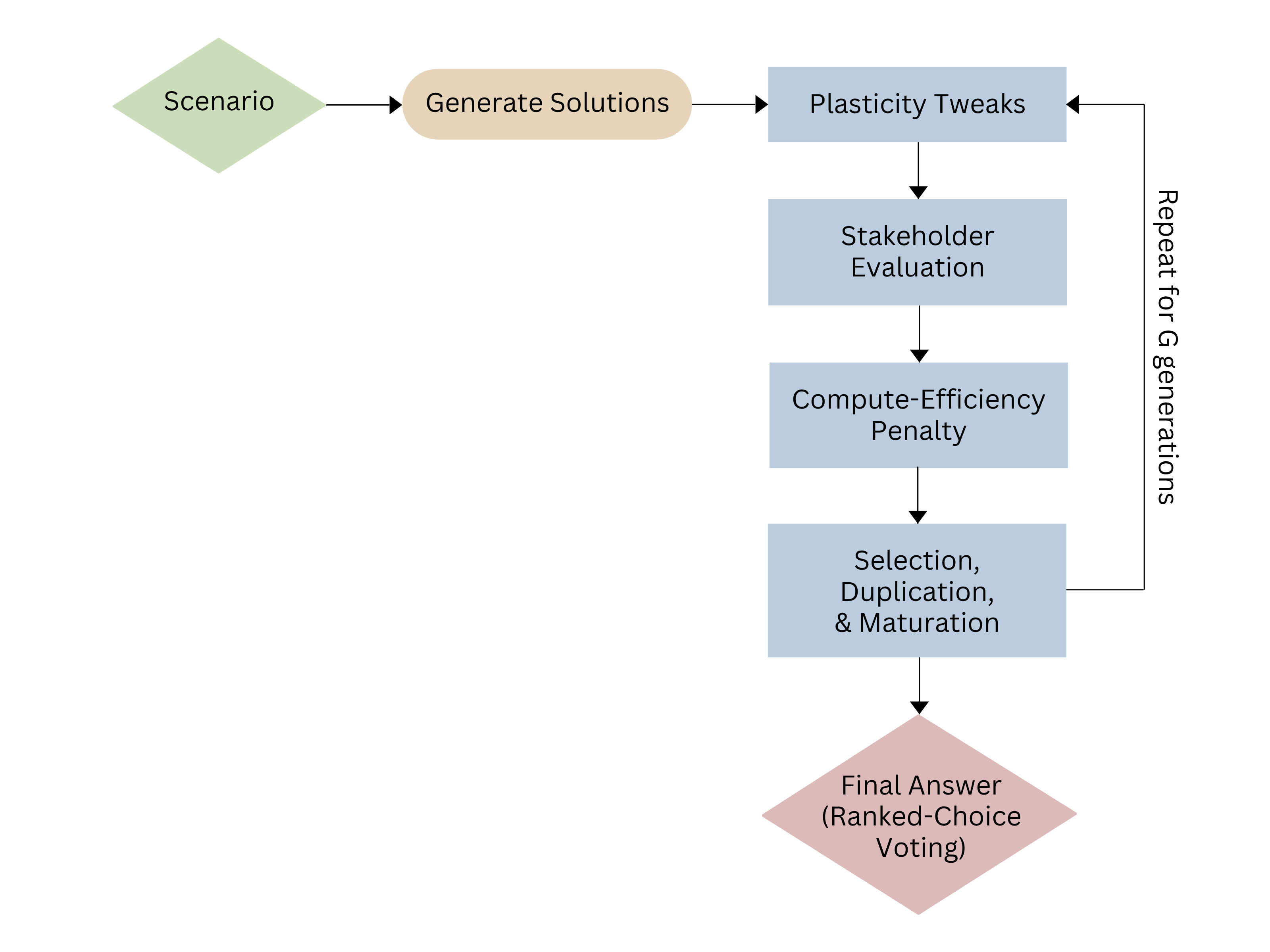}
\caption{The workflow of the Lark evolutionary framework with the four core mechanisms: plasticity tweaks, stakeholder evaluation, and duplication/maturation with compute-efficiency penalties across generations.}
\label{fig:lark_workflow}
\end{figure}

We introduce Lark, a biology-inspired neuroevolutionary framework that couples LLM-driven reasoning with discrete-generation evolutionary search for multi-stakeholder decision-making. Unlike reinforcement learning approaches that require Markov decision process (MDP) formulations and continuous state-action transitions~\citep{Sutton2018}, Lark operates in a discrete generational paradigm where populations of candidate strategies evolve through selection, mutation, and specialization~\citep{Stanley2002, Surina2025}. This design choice is motivated by three key considerations: (1) multi-stakeholder strategy generation lacks the sequential state transitions and immediate reward signals needed for effective RL, (2) evolutionary algorithms excel at exploring large discrete solution spaces without gradient information~\citep{Cuccu2018, Miikkulainen2019}, and (3) LLM-based generation naturally aligns with population-based search where diverse proposals are evaluated holistically rather than incrementally~\citep{Surina2025, Chen2025}.

Inspired by AlphaEvolve~\citep{Novikov2025}, which demonstrated that evolutionary refinement can solve complex problems through iterative population improvement, we extend this paradigm by integrating four biologically motivated mechanisms: (1) \textbf{plasticity} for within-generation context-sensitive refinement, (2) \textbf{duplication and maturation} for modular specialization, (3) \textbf{ranked-choice stakeholder voting} for preference aggregation without cardinal utility assumptions, and (4) \textbf{compute-awareness} via token-based penalties that reward brevity. These mechanisms mirror natural evolutionary processes—synaptic adaptation, gene duplication with subfunctionalization, collective decision-making, and resource constraints—that have proven effective across biological systems~\citep{Soltoggio2008, Clune2013, Black1986}. Figure~\ref{fig:lark_workflow} provides an overview of the complete Lark evolutionary loop.

\subsection{Problem Setting}

Let $\mathcal{X}$ denote the space of natural language strategy descriptions. Given a decision-making scenario $\mathcal{D} = (\mathcal{C}, \mathcal{O}, \mathcal{S})$ consisting of context $\mathcal{C}$, objectives $\mathcal{O}$, and $m$ stakeholders $\mathcal{S} = \{s_1, \ldots, s_m\}$ with heterogeneous preferences, the goal is to generate a set of strategies $\mathbf{x} = \{x_1, \ldots, x_k\} \subset \mathcal{X}$ that collectively satisfy stakeholder preferences while respecting a compute budget $C_{\max}$. Each stakeholder $s_j$ provides ordinal rankings over candidate strategies based on their utility function $u_j: \mathcal{X} \to \mathbb{R}$. We aggregate these rankings using influence-weighted Borda scoring (Section 3.4) to produce consensus scores $B(x_i)$, which guide evolutionary selection. Unlike MDP-based formulations that require state-action-reward dynamics, this discrete optimization setting evaluates entire strategy trajectories holistically, making evolutionary search more suitable than temporal-difference methods~\citep{Salimans2017, Such2017}.

\subsection{Plasticity: Context-Sensitive Refinement}

Inspired by Hebbian learning and neuromodulation in biological systems~\citep{Soltoggio2008}, plasticity enables agents to make concise, context-aware adjustments to candidate strategies within a generation. Rather than modeling continuous weight dynamics (which requires differentiable objectives), we implement plasticity as an LLM-driven refinement operator $\Phi: \mathcal{X} \times \mathcal{C} \to \mathcal{X}$ that takes a candidate strategy $x$ and scenario context $\mathcal{C}$ to produce an improved variant $x' = \Phi(x, \mathcal{C})$.

The plasticity prompt instructs the LLM to:
\begin{itemize}
\item Identify context-specific weaknesses in the current strategy
\item Propose targeted modifications addressing these weaknesses
\item Maintain the core structure while refining implementation details
\item Limit modifications to preserve solution diversity (brevity constraint)
\end{itemize}

This mechanism accelerates convergence by enabling rapid within-generation adaptation without requiring full strategy regeneration~\citep{Soltoggio2008}. We apply plasticity with probability $p_{\text{plast}}$ after initial strategy generation, controlled by a temperature parameter that decays across generations to balance exploration and exploitation.

\subsection{Duplication and Maturation: Modular Specialization}

Motivated by gene duplication and subfunctionalization in evolutionary biology~\citep{Clune2013, Stanley2002}, high-performing strategies are probabilistically duplicated and specialized to promote functional diversity. Let $R(x_i)$ denote the aggregated stakeholder score for strategy $x_i$ (derived from Borda counts in Section 3.4). The duplication probability follows fitness-proportional selection:
\begin{equation}
p_{\text{dup}}(x_i) = \sigma\left(\frac{R(x_i) - \bar{R}}{\tau}\right),
\end{equation}
where $\sigma(z) = (1 + e^{-z})^{-1}$ is the logistic function, $\bar{R} = \frac{1}{k}\sum_{i=1}^k R(x_i)$ is the mean population fitness, and $\tau > 0$ is a temperature parameter controlling selection pressure.

Duplicated strategies undergo \textbf{maturation}, a stochastic specialization process where the LLM receives a prompt to:
\begin{itemize}
\item Identify a specific stakeholder subgroup or objective dimension
\item Modify the duplicated strategy to specialize for that subgroup
\item Introduce novel implementation details not present in the parent
\item Ensure the variant remains feasible within stated constraints
\end{itemize}

This models the subfunctionalization observed in gene duplication events, where duplicated genes diverge to perform distinct but related functions~\citep{Clune2013}. Unlike traditional genetic operators that apply random mutations, maturation leverages LLM semantic understanding to generate meaningful specializations~\citep{Surina2025, Chen2025}.

\subsection{Stakeholder Ranked-Choice Voting}

To aggregate heterogeneous stakeholder preferences without imposing cardinal utility assumptions or eliciting interpersonal comparisons, we employ influence-weighted Borda count~\citep{Black1986}. Each stakeholder $s_j$ provides a ranking $\pi_j$ over the $k$ candidate strategies. The Borda score for strategy $x_i$ is:
\begin{equation}
B(x_i) = \sum_{j=1}^m w_j \cdot (k - \text{rank}_{\pi_j}(x_i)),
\end{equation}
where $\text{rank}_{\pi_j}(x_i) \in \{1, \ldots, k\}$ denotes the position of $x_i$ in stakeholder $j$'s ranking, and $w_j \geq 0$ is stakeholder $j$'s influence weight (with $\sum_{j=1}^m w_j = 1$). The consensus strategy is $\hat{x} = \arg\max_i B(x_i)$.

To quantify preference alignment across stakeholders, we compute the coefficient of variation:
\begin{equation}
\text{CV}(B) = \frac{\sqrt{\frac{1}{k}\sum_{i=1}^k (B(x_i) - \bar{B})^2}}{\bar{B}},
\end{equation}
where $\bar{B} = \frac{1}{k}\sum_{i=1}^k B(x_i)$. Lower CV indicates stronger consensus. This mechanism is robust to strategic manipulation and does not require stakeholders to reveal cardinal utilities~\citep{Black1986}.

\subsection{Compute-Efficiency Penalty}

To enforce resource-aware optimization and reward brevity, we incorporate a token-based penalty into fitness evaluation. Let $T(x_i)$ denote the token count of strategy $x_i$. The compute-adjusted fitness is:
\begin{equation}
R(x_i) = B(x_i) \cdot \left(1 - \lambda \cdot \max\left(0, \frac{T(x_i) - T_{\text{target}}}{T_{\text{target}}}\right)\right),
\end{equation}
where $T_{\text{target}}$ is a target token budget and $\lambda \in [0, 1]$ is a penalty coefficient. This soft constraint encourages compact, efficient strategies without imposing hard limits that may exclude high-quality verbose solutions. We track a running efficiency metric:
\begin{equation}
E^{(t)} = \frac{1}{k} \sum_{i=1}^k \frac{B(x_i)}{T(x_i)},
\end{equation}
which measures the average quality-per-token across the population at generation $t$. Monitoring $E^{(t)}$ reveals whether the system improves efficiency over time.

\subsection{Evolutionary Loop}

Algorithm~\ref{alg:lark} integrates all mechanisms into a unified discrete-generation evolutionary framework. Unlike RL methods that update policies after each action~\citep{Sutton2018}, Lark evaluates entire strategy populations holistically at each generation, enabling global comparison and selection~\citep{Salimans2017, Such2017}. This design is particularly suited for LLM-based search where generation is discrete and expensive, making population-based evaluation more sample-efficient than sequential trial-and-error~\citep{Surina2025}.

\begin{algorithm}[H]
\caption{Lark Evolutionary Loop}
\label{alg:lark}
\begin{algorithmic}[1]
\STATE \textbf{Input:} Scenario $\mathcal{D} = (\mathcal{C}, \mathcal{O}, \mathcal{S})$, generations $G$, population size $k$, target tokens $T_{\text{target}}$
\STATE \textbf{Output:} Optimized strategy set and efficiency trajectory $\{E^{(t)}\}_{t=1}^G$
\STATE Initialize population $\mathcal{P}_0 = \{x_1^{(0)}, \ldots, x_k^{(0)}\}$ via LLM sampling conditioned on $\mathcal{D}$
\FOR{$t = 1$ to $G$}
    \STATE \textbf{Plasticity:} For each $x_i \in \mathcal{P}_{t-1}$, apply $\Phi(x_i, \mathcal{C})$ with probability $p_{\text{plast}}$
    \STATE \textbf{Stakeholder Evaluation:} Each stakeholder $s_j$ ranks all strategies $\to$ rankings $\{\pi_j\}_{j=1}^m$
    \STATE \textbf{Scoring:} Compute influence-weighted Borda scores $B(x_i)$ via Eq.~(2)
    \STATE \textbf{Compute Penalty:} Adjust scores via token penalty: $R(x_i) \leftarrow B(x_i) \cdot (1 - \lambda \cdot \max(0, (T(x_i) - T_{\text{target}})/T_{\text{target}}))$
    \STATE \textbf{Selection:} Compute duplication probabilities $p_{\text{dup}}(x_i)$ via Eq.~(1)
    \STATE \textbf{Duplication:} Sample strategies for duplication according to $p_{\text{dup}}$
    \STATE \textbf{Maturation:} Apply LLM-driven specialization to duplicated strategies
    \STATE Construct next generation $\mathcal{P}_t$ by selecting top-$k$ strategies from $\mathcal{P}_{t-1} \cup \{\text{duplicated strategies}\}$
    \STATE Track efficiency: $E^{(t)} \leftarrow \frac{1}{k}\sum_{i=1}^k B(x_i)/T(x_i)$
\ENDFOR
\STATE \textbf{return} $\mathcal{P}_G$ and efficiency trajectory $\{E^{(t)}\}_{t=1}^G$
\end{algorithmic}
\end{algorithm}

\textbf{Justification for Evolutionary vs. RL Approach.} We adopt discrete-generation evolutionary search over continuous RL for four reasons grounded in recent literature~\citep{Salimans2017, Such2017, Surina2025}: (1) \textbf{No MDP Structure}: Multi-stakeholder strategy generation lacks sequential state transitions and immediate rewards—strategies are evaluated holistically after completion, not incrementally. (2) \textbf{Sparse Learning Signal}: Stakeholder preferences provide sparse ordinal feedback (rankings), which is insufficient for gradient-based policy updates but well-suited for population-level selection. (3) \textbf{Exploration Efficiency}: Evolutionary methods excel at exploring large discrete spaces (natural language strategies) where gradients are unavailable or misleading~\citep{Cuccu2018}. (4) \textbf{LLM Integration}: Recent work on LLM-driven evolutionary search demonstrates superior performance over RL in algorithm discovery tasks, as generation naturally aligns with population sampling~\citep{Surina2025, Chen2025}. This positions Lark as a practical, compute-aware evolutionary system rather than a formal constrained optimization framework.

\section{Experiments}



\subsection{Experimental Setup}

This study presents \textbf{preliminary results} from an initial investigation of Lark's mechanisms using controlled synthetic scenarios. As a proof-of-concept exploration, these experiments establish baseline feasibility and inform the design of more comprehensive evaluations planned for future work.

\subsubsection{Evaluation Protocol}
We evaluated Lark using an LLM-as-a-judge framework with dual independent judges to mitigate single-evaluator biases. Each judge assessed model outputs using a 50-point rubric with five equally weighted criteria (10 points each): (1) Coverage/Completeness (addresses all key stakeholder concerns), (2) Feasibility/Realism (practical implementability), (3) Specificity/Thoroughness (concrete operational details), (4) Constraint Adherence (respects stated limitations), and (5) Clarity/Structure (logical organization). We used \texttt{openai/gpt-oss-120b}\footnote{An open-source GPT-based model selected for inference efficiency under computational constraints.} as the judge model. To reduce bias, system identities were \textbf{blinded} (anonymized IDs) and, for each scenario, the \textbf{input order of systems was randomized} per judge. Judge aggregation employed ranked-choice voting with temperature = 0.1 for consistency.

\subsubsection{Benchmark Design}
We constructed 30 unique synthetic scenarios spanning six decision-making domains: Multi-Stakeholder Trade-offs (5 scenarios), Policy Proposal (5), Product Roadmap (5), Campaign Plan (5), Infrastructure Siting (5), and Clinical Decision-Making (5). Each scenario was synthetically generated with controlled complexity using prompt templates that specified stakeholder profiles, constraints, and conflicting objectives.

\textbf{Rationale for Synthetic Benchmarks.} We deliberately use synthetic scenarios for this preliminary investigation to enable controlled experimentation and systematic ablation studies. This approach allows us to isolate the contribution of each mechanism while maintaining experimental control over difficulty levels and stakeholder complexity. We acknowledge that real-world validation is essential for assessing ecological validity and plan to incorporate authentic multi-stakeholder case studies in subsequent phases of this research.

\subsubsection{Model Comparison}
We benchmarked Lark (implemented with DeepSeek-V3.1 as the base model) against nine contemporary LLMs: DeepSeek-V3.1 (baseline), Qwen3-Next-80B-A3B-Thinking, GPT-5-nano, GPT-5-mini, GPT-4o, GPT-4.1, o4-mini, o3-mini, and o3. Due to funding limitations for this preliminary study, Claude and other proprietary families were excluded.

\subsubsection{Ablation Studies}
To isolate mechanism contributions, we evaluated four ablation variants: (1) \textbf{Lark-NoPlasticity} (disables context-sensitive refinement), (2) \textbf{Lark-NoRankedChoiceVoting} (replaces Borda count with simple averaging), (3) \textbf{Lark-NoMutationAndNoDuplication} (removes modular specialization mechanisms), and (4) \textbf{Lark-NoPenalty} (eliminates compute-awareness constraints). Full Lark results represent performance after five evolutionary generations. All variants were evaluated on identical scenarios under matched conditions.

\subsubsection{Statistical Analysis}
We report per-model means with \textbf{95\% CIs} across the 30 rounds. Pairwise comparisons vs Lark–Full use \textbf{paired Wilcoxon signed-rank} tests (two-sided, \(\alpha=0.05\)). Effect sizes report \textbf{Cohen’s \(d_z\)} from paired differences. Costs are reported as provided means (per-round cost logs were not retained).


\section{Results and Analysis}
\paragraph{Setup and unit of analysis.}
We ran \textbf{30 rounds} per model. All statistics treat rounds as the unit of analysis (\(n{=}30\)). For each round, we compute model \emph{rank} (1=best) from the composite rubric score (/50) across the 14 systems.

\subsection{Overall performance}
Across 30 rounds, \textbf{Lark Full} achieves the best mean position and the highest composite quality: \textbf{mean rank} \(2.55\) (95\% CI \([2.17, 2.93]\)) and \textbf{mean score} \(29.4/50\) (95\% CI \([26.34, 32.46]\)). Strong baselines trail (e.g., GPT-o3 mean rank \(4.30\), score \(28.8/50\); Qwen3-Next-80B mean rank \(4.40\), score \(27.9/50\)).

\begin{table}[H]
\centering
\caption{Overall quality, rank, and cost (30 rounds). Mean rank and mean composite score include 95\% confidence intervals. Costs are per task as provided.}
\label{tab:overall}
\small
\begin{tabular}{l
                S[table-format=2.2]
                l
                S[table-format=2.1]
                l
                S[table-format=1.6]}
\toprule
\textbf{Model} &
\multicolumn{2}{c}{\textbf{Mean Rank} $\downarrow$} &
\multicolumn{2}{c}{\textbf{Mean Score /50} $\uparrow$} &
\textbf{Avg.\ Cost (\$/task)} \\
\cmidrule(lr){2-3}\cmidrule(lr){4-5}
& {Value} & {95\% CI} & {Value} & {95\% CI} &  \\
\midrule
\textbf{Lark Full} &
\bfseries 2.55 & \bfseries [2.17, 2.93] &
\bfseries 29.4 & \bfseries [26.34, 32.46] &
\bfseries 0.016006 \\

GPT o3 & 4.30 & [3.10, 5.50] & 28.8 & [26.30, 31.30] & 0.016424 \\
Qwen3 Next 80B A3B Thinking & 4.40 & [2.98, 5.82] & 27.9 & [23.99, 31.81] & 0.005729 \\
Lark NoPenalty & 6.15 & [5.41, 6.89] & 27.2 & [24.28, 30.12] & 0.011163 \\
GPT o4-mini & 6.50 & [5.44, 7.56] & 27.2 & [24.64, 29.76] & 0.008800 \\
Lark NoRankedChoiceVoting & 6.55 & [5.30, 7.80] & 27.0 & [21.69, 32.31] & 0.006070 \\
GPT 4.1 & 7.25 & [5.61, 8.89] & 26.9 & [23.95, 29.85] & 0.013640 \\
Lark NoMutationNoDuplication & 8.95 & [7.90, 10.00] & 25.9 & [22.50, 29.30] & 0.006053 \\
Lark NoPlasticity & 8.30 & [7.24, 9.36] & 26.2 & [22.70, 29.70] & 0.010323 \\
GPT o3-mini & 9.60 & [8.40, 10.80] & 25.6 & [22.67, 28.53] & 0.010217 \\
GPT 4o & 9.50 & [7.17, 11.83] & 25.4 & [18.60, 32.20] & 0.013270 \\
GPT5 Mini & 8.60 & [5.67, 11.53] & 25.4 & [20.00, 30.80] & 0.006750 \\
Deepseek V3.1 (aka Lark Base) & 10.60 & [8.61, 12.59] & 23.7 & [15.60, 31.80] & 0.001568 \\
GPT5 Nano & 11.75 & [10.01, 13.49] & 23.0 & [18.00, 28.00] & 0.001940 \\
\bottomrule
\end{tabular}
\end{table}
\FloatBarrier

\subsection{Mechanism ablations}
We ablate four levers---\emph{Plasticity}, \emph{Duplication+Maturation}, \emph{Ranked aggregation (RCV)}, and the \emph{Token Penalty}---and report deltas relative to Lark Full. Positive \(\Delta\)Score favors Lark Full. Positive \(\Delta\)Rank means the variant ranks worse.

\begin{table}[H]
\centering
\caption{Ablation deltas vs.\ Lark Full across 30 rounds. \(\Delta\)Score \(=\) Score\(_\mathrm{Full}\)\(-\)Score\(_\mathrm{Variant}\); \(\Delta\)Rank \(=\) Rank\(_\mathrm{Variant}\)\(-\)Rank\(_\mathrm{Full}\). 95\% CIs by normal approximation over rounds.}
\label{tab:ablations}
\small
\begin{tabular}{l S[table-format=2.1] l S[table-format=2.2] l}
\toprule
\textbf{Variant} & {\(\Delta\)Score (/50) $\uparrow$} & \textbf{95\% CI} & {\(\Delta\)Rank (pos.) $\downarrow$} & \textbf{95\% CI} \\
\midrule
Lark\,—\,Penalty Off & 2.2 & [1.72, 2.68] & 3.60 & [2.89, 4.31] \\
Lark\,—\,RCV Off     & 2.4 & [1.69, 3.11] & 4.00 & [2.82, 5.18] \\
Lark\,—\,Plasticity Off & 3.4 & [2.74, 4.06] & 5.75 & [4.72, 6.78] \\
Lark\,—\,No Mutation \& No Duplication & 3.5 & [3.00, 4.00] & 6.40 & [5.53, 7.27] \\
\bottomrule
\end{tabular}
\end{table}
\FloatBarrier

\subsection{Pairwise significance tests and effect sizes}
We perform \emph{paired Wilcoxon signed-rank tests} (two-sided, \(\alpha{=}0.05\)) on per-round composite scores (\(n{=}30\)), comparing Lark Full against ablations and strong baselines. We also report paired \emph{Cohen’s \(d_z\)}.

\begin{table}[H]
\centering
\caption{Lark Full vs.\ comparators on composite scores (30 rounds). \(\Delta\)Mean is Full minus comparator (/50). Cohen’s \(d_z\) from paired differences.}
\label{tab:wilcoxon}
\small
\begin{tabular}{l S[table-format=2.1] S[table-format=1.2] l}
\toprule
\textbf{Comparator} & {\(\Delta\)Mean (/50)} & {\(d_z\)} & \textbf{Wilcoxon \(p\)} \\
\midrule
Lark NoPenalty & 2.2 & 1.63 & 5.93e-06 \\
Lark NoRankedChoiceVoting & 2.4 & 1.20 & 1.94e-05 \\
Lark NoPlasticity & 3.4 & 1.86 & 5.93e-06 \\
Lark NoMutationAndNoDuplication & 3.5 & 2.53 & 1.83e-06 \\
\midrule
GPT o3 & 0.6 & 0.20 & 0.290 \\
Qwen3 Next 80B A3B Thinking & 1.5 & 0.28 & 0.548 \\
GPT 4.1 & 2.5 & 0.90 & 1.40e-04 \\
GPT 4o & 4.0 & 2.54 & 1.83e-06 \\
Deepseek V3.1 (aka Lark Base) & 5.7 & 1.12 & 1.83e-06 \\
\bottomrule
\end{tabular}
\end{table}
\FloatBarrier

\subsection{Costs}
Average cost per task for the top two systems is comparable: Lark Full \(\$0.0160\) vs.\ GPT--o3 \(\$0.0164\). Comprehensive cost figures for all systems are included in Table~\ref{tab:overall}. \textit{We did not enforce token caps; reported costs reflect natural output lengths under model defaults.}

\section{Discussion and Broader Impacts}

Lark integrates four complementary mechanisms---\emph{plasticity}, \emph{duplication+maturation}, \emph{stakeholder-ranked aggregation}, and a \emph{compute-efficiency penalty}---to address verbosity, narrow exploration, and weak multi-stakeholder handling in LLM multi-agent systems.

\subsection{Discussion}

\paragraph{Summary of findings.}
Across 30 rounds, \textbf{Lark Full} is competitive with strong proprietary models and clearly outperforms its ablations. Gains are both \emph{statistically reliable} (paired Wilcoxon \(p<0.05\) for all ablation contrasts) and \emph{practically meaningful}: a \(+2\)–\(3.5\) point lift on a 50-point rubric corresponds to \(\sim\)4–7\% of the scale and a mean-rank improvement of \(+3.6\)–\(6.4\) positions (Table~\ref{tab:ablations}). Differences vs.\ GPT-o3 and Qwen-Next-80B are small and non-significant (parity on these rounds), while improvements vs.\ GPT-4.1 and GPT-4o are large to very large (Table~\ref{tab:wilcoxon}).

\paragraph{Cost--quality trade-off.}
Lark Full is cost-comparable to GPT-o3 (\$0.0160 vs.\ \$0.0164 per task; Table~\ref{tab:overall}). \emph{No token caps were enforced;} costs reflect natural output lengths under model defaults. Ablations are cheaper but give up \(2\)–\(3.5\) points of quality, placing Lark Full near a Pareto frontier for these tasks.

\subsection{Mechanism-level insights}
\begin{itemize}
  \item \textbf{Duplication+maturation} and \textbf{plasticity} are the principal drivers (\(\Delta\)Score \(=3.5\) and \(3.4\); \(d_z=2.53\) and \(1.86\)), indicating that within-generation refinement and specialization materially improve solution quality.
  \item \textbf{Ranked aggregation (RCV)} provides substantive gains (\(\Delta\)Score \(=2.4\); \(d_z=1.20\)), suggesting stakeholder-aware selection outperforms simple averaging for consensus utility.
  \item The \textbf{compute-efficiency penalty} reduces verbosity without harming quality (\(\Delta\)Score \(=2.2\) vs.\ no-penalty), aligning with concise, compute-aware reasoning.
\end{itemize}

\subsection{Interpreting Wilcoxon \(p\) and Cohen’s \(d_z\)}
The paired \emph{Wilcoxon signed-rank} test evaluates per-round differences \(d_i=\text{score}_\text{Full}^{(i)}-\text{score}_\text{comp}^{(i)}\) and asks whether the \emph{median} \(d_i\) is zero; it does not assume normality (ties are omitted). We adjust \(p\)-values using \emph{Holm}'s step-down procedure; claims of significance refer to adjusted values. \emph{Cohen’s \(d_z\)} quantifies the \emph{magnitude} of the paired effect (\(d_z=\bar d/s_d\)) with conventional guidelines: 0.2\,=\,small, 0.5\,=\,medium, 0.8\,=\,large, \(>\)1.2\,=\,very large. Our ablation contrasts are large to very large; GPT-4.1 is large; GPT-o3 and Qwen are small and non-significant on these rounds. For context, \(\Delta\)Mean \(+2\)–\(3.5\) (of 50) translates to \(\sim\)4–7\% of scale and rank shifts of \(+3.6\)–\(6.4\) places.

\subsection{Scope and expected generalization}
We expect the largest relative gains in settings that require (i) proposing \emph{diverse futures} and then \emph{specializing} them for constituencies, and (ii) explicit \emph{multi-stakeholder trade-offs}. Benefits should be smaller on single-objective, tightly constrained tasks where exploration and negotiation matter less.

\subsection{Limitations and future directions}
This study demonstrates \emph{proof-of-concept} feasibility but leaves open: (1) ecological validity on real multi-stakeholder problems; (2) broader model coverage and hyperparameter sweeps; (3) stability of ranked aggregation with larger stakeholder sets; (4) direct measurements of wall-clock time and energy. We plan: (i) real-world validation in policy/healthcare/organizational settings; (ii) compute/energy audits and comparisons to multi-objective baselines; (iii) scalability tests with 10--50 stakeholders and richer preference structures; (iv) expanded baselines (including specialized multi-agent frameworks); and (v) longitudinal deployment to assess iterative decision cycles.

\paragraph{Broader impacts.}
Compute-aware design can reduce environmental cost by discouraging verbosity; stakeholder-aware aggregation can promote more equitable outcomes across constituencies. Ethical deployment requires transparency around aggregation choices, compute budgets, and failure modes.

\section{Conclusion}
We present Lark, a biologically inspired evolutionary framework for multi-agent LLM systems. Plasticity, duplication, stakeholder voting, and compute-efficient optimization enable rapid adaptation, multi-stakeholder fairness, and modular architectures. Preliminary results demonstrate promising improvements over baselines. Future work includes richer multi-agent domains, larger LLM integration, and real-world deployment.

\newpage
\bibliographystyle{plainnat}

\end{document}